\documentclass[aip, amsmath, amssymb, reprint]{revtex4-2}
\usepackage{graphicx}
\usepackage{dcolumn}
\usepackage{bm}

\usepackage[utf8]{inputenc}
\usepackage[T1]{fontenc}
\usepackage{etoolbox}
\usepackage{natbib}

\makeatletter
\def\@email#1#2{%
	\endgroup
	\patchcmd{\titleblock@produce}
	{\frontmatter@RRAPformat}
	{\frontmatter@RRAPformat{\produce@RRAP{*#1\href{mailto:#2}{#2}}}\frontmatter@RRAPformat}
	{}{}
}%
\makeatother

\begin{document}	
\title{First-principles prediction of structural, magnetic properties of Cr-substituted strontium hexaferrite, and its site preference}
\author{Binod Regmi, ${ }^{1,2}$ Dinesh Thapa, ${ }^{1,2}$ 
Bipin Lamichhane, ${ }^{1,2}$ Seong-Gon Kim, ${ }^{1,2, a)}$ \\
${ }^{1}$ Department of Physics and Astronomy, Mississippi State University, Mississippi State, Mississippi 39762, USA \\
${ }^{2}$ Center for Computational Sciences, Mississippi State University, Mississippi State, Mississippi 39762, USA} 
\altaffiliation {Author to whom correspondence should be addressed. Electronic mail: sk162@msstate.edu}
	
\begin{abstract}
	ABSTRACT\\	
	To investigate the structural and magnetic properties of Cr-doped M-type strontium hexaferrite (SrFe$_{12}$O$_{19}$) with x = (0.0, 0.5, 1.0), we perform first-principles total-energy calculations relied on density functional theory. Based on the calculation of the substitution energy of Cr in strontium hexaferrite and formation probability analysis, we conclude that the doped Cr atoms prefer to occupy the 2a, 12k, and 4f$_{2}$ sites which is in good agreement with the experimental findings. Due to Cr$^{3+}$ ion moment, 3 {$\mu_B$}, smaller than that of Fe$^{3+}$ ion, 5 {$\mu_B$}, saturation magnetization (M$_{s}$) reduce rapidly as the concentration of Cr increases in strontium hexaferrite. The magnetic anisotropic field $\left(H_{a}\right)$ rises with an increasing fraction of Cr despite a significant reduction of magnetization and a slight increase of magnetocrystalline anisotropy $\left(K_{1}\right)$.The cause for the rise in magnetic anisotropy field $\left(H_{a}\right)$ with an increasing fraction of Cr is further emphasized by our formation probability study. Cr$^{3+}$ ions prefer to occupy the 2a sites at lower temperatures, but as the temperature rises, it is more likely that they will occupy the 12k site. Cr$^{3+}$ ions are more likely to occupy the 12k site than the 2a site at a specific annealing temperature (>700°C).
\end{abstract}	
\maketitle
\section{\label{sec:level1} Introduction}
\indent Hexaferrites, also known as hexagonal ferrites or hexagonal ferrimagnets, are a class of magnetic materials that have been of great interest to researchers since their discovery in the $1950$’s. These hexaferrites are found in numerous types such as M, Y, Z, W, X, U-type commonly doped with zinc, strontium, nickel, aluminum, and magnesium. The most common properties to all hexaferrites include that all are ferrimagnetic, their properties of magnetism are based on the crystal structure, and they take different amounts of energy to magnetize in a specific direction within the crystal because of spin-orbit interaction \cite{pullar2012hexagonal}.\\ 
\indent Particularly, we are interested in M-type strontium hexaferrite (SrFe$_{12}$O$_{19}$, SFO) that falls to space group $P63/mmc$ which has a crystal structure of hexagonal magnetoplumbite. The unit cell of SFO having two formula units is presented in Fig(1). 
\begin{figure}[hb]
         \includegraphics[width=1.0\linewidth]{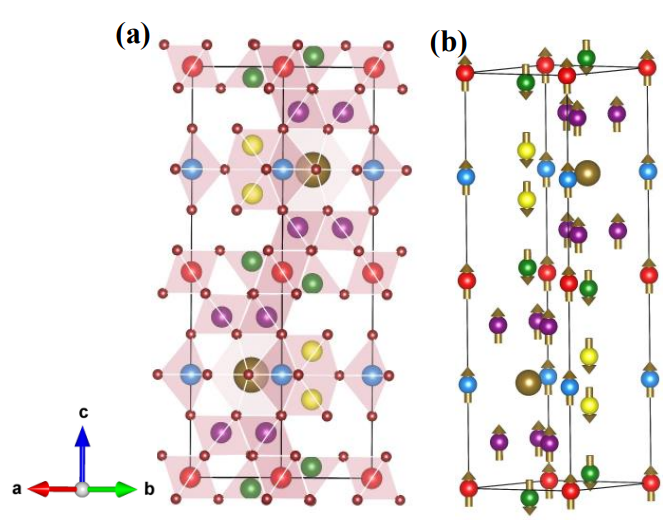}
         \caption{(a) One double formula unit cell of $\mathrm{SrFe}_{12}\mathrm{O}_{19}$. Small maroon spheres are $\mathrm {O}$ atoms, and two huge gray spheres are Sr atoms. $\mathrm{Fe}^{3+}$ ions are represented by colored spheres encircled by polyhedra made of $\mathrm{O}$ atoms in a variety of inequivalent sites: $2a$ (red), $2b$ (blue), $4f_{1}$ (green), $4f_{2}$ (yellow), and $12 k$ (purple).(b) A schematic representation of the $\mathrm{Fe}^{3+}$ ions of $\mathrm{SrFe}_{12}\mathrm{O}_{19}$ in their lowest energy spin configuration. The local magnetic moment at each atomic location is represented by the arrows.}
\end{figure}
	The iron ions in this structure are coupled in a tetrahedral, trigonal bipyramidal, and octahedral manner by oxygen ions.The magnetic property is retained in SFO mainly due to the occupancy of Fe$^{+3}$ ions in five inequivalent sites (namely 2a, 2b, 4f$_1$, 4f$_2$, and 12k), three octahedral sites (namely 2a, 12k, and 4f$_2$), one trigonal bipyramidal site (2b), and one tetrahedral site (4f$_1$). However, the range (degree) of magnetic properties will be influenced by the shape and size of material particles mainly in the context of thin films, nanoparticles \cite{dixit2017effect,fang2003magnetic, soria2019strontium,obradors1988crystal}. \\
\indent In SFO, there is an involvement of interactions between the moments or moments with lattice ions, tend to contribute anisotropic energy which is termed as magnetocrystalline anisotropy (MA). In a more explicit way, it is the dependence of the magnetic properties on the applied field direction relative to the crystal lattice. Magnetocrystalline anisotropy energy (MAE), an integral property of a ferromagnetic crystal, is the energy difference to magnetize a crystal along easy and hard direction of magnetization \cite{dixit2017effect, Aharoni1996}. The primary source of MA is spin-orbit coupling (SOC). Because of this coupling, orbitals of electrons are coupled with the spin of electrons and follow the spin direction no matter how the magnetization changes its direction in space \cite{Aharoni1996}. The anisotropy that arises on a crystal are mainly due to the shape of a magnetic particle at the quantum scale, atomic diffusion at sufficiently high temperature, and the interaction between a ferromagnetic and an antiferromagnetic materials \cite{Aharoni1996, miyazaki2012physics,meiklejohn1956new,hong2019influencing}.\\
\indent SFO is one of the best candidates among hexaferrite groups owing to its industrial and electronic implications. In the beginning years, Technophiles were motivated about SFO to make permanent magnets, recording media, electric motors because of its high saturation magnetization, large coercivity, optimum Curie temperature, supreme magnetocrystalline anisotropy, and more chemical stability. Nowadays, due to technological advancement, it is equally growing interest in the development of nano fibres, electronic components for mobile and wireless communications. Recently, researchers have been characterized Y, M, U, and Z ferrites as multiferroics even at room temperature. These multiferroics have a wide range of practical implications like multi-state memory elements, memory media, and novel functional sensors \cite{dixit2017effect,li2016magnetic,tan2013structure}.\\
\indent Several successful investigations have been done on SFO to understand its electronic structure by computational and experimental approach. To uplift the strength of magnetic and electric properties, researchers are substituted ions or pair of ions in a different concentration mainly on Fe sites of SFO. Majority of researchers have followed the non-magnetic ions substitution into Fe sites to enhance further the value of saturation magnetization (M$_s$). In case of Zr-Cd substitution (SrFe$_{12-2x}$(ZrCd)$_{x}$O$_{19}$), the value of M$_s$ augmented in the limit of concentration $x = 0.2$, whereas the value of coercivity declined with increasing concentration of Zr-Cd \cite{ashiq2009structural}. Substitution of Er-Ni pair in SFO showed the continuous rise in the value of M$_s$ and coercivity in accordance with the concentration \cite{ashiq2012synthesis}. However, the substitution of certain pairs like Zn-Nb,\cite{fang2004temperature} Zn-Sn, \cite{ghasemi2010role,ghasemi2011correlation,dixit2019site} and Sn-Mg \cite{davoodi2011magnetic,ghasemi2009microwave} showed the increasing trend of M$_s$ and decreasing pattern of coercivity. \\
\indent In this study, we performed the first-principles total-energy calculations to analyze the link between site occupation and magnetic properties of substituted strontium hexaferrite, SrFe$_{12-x}$Cr$_{x}$O$_{19}$ with $x = 0.5$ and $x = 1.0$. Every configuration of substituted SFO appears with a particular probability. To determine the formation probabilities of its various configurations at a typical annealing temperature ($1000 K$), we used the Boltzmann distribution function. We show that our calculation predicts a decrease of saturation magnetization ($M_{s}$) as well as a decrease in magnetic anisotropy energy (MAE) of SrFe$_{12-x}$(Cr)$_{x}$O$_{19}$ at $x = 0.5$ and $1.0$ compared to the pure M-type SFO. This result is also in good agreement with the experimental observation as observed in Ghasemi et al. (2009) \cite{ghasemi2009microwave}.
\section{\label{sec:level2}Computational details}
\indent We pursued the first-principles total-energy calculations for configuration SrFe$_{12-x}$(Cr)$_x$O$_{19}$ at $x = 0.5$ and $1.0$. In our calculation, a unit cell of two formula	units of SFO is used. The structural optimization calculation along with total energies and forces were carried out using density functional theory by projector augmented wave (PAW) potential as executed in VASP. Depending on the ground state ferrimagnetic spin ordering of Fe, our calculations were solely focused on spin-polarized \cite{fang2003magnetic, gorter1957saturation}. The expansion of the wave function was in the form of plane waves with a $520 eV$ energy cut-off used for pristine SFO, Cr-substituted SFO. A $7 \times 7 \times 1$ Monkhorst-Pack k-mesh was used to sample a Brillouin zone with a Fermi-level smearing of $0.2 eV$ applied through the Methfessel-Paxton method \cite{monkhorst1976special,methfessel1989high}. The electronic relaxation was accomplished till the change in free energy and the band structure energy less than $10^{-7} eV$. In addition, we fully optimized the structure by relaxing the positions of ions and cell shape till the change in total energies between two ionic steps less than $10^{-4} eV$. We used the Perdew-Burke-Ernzerhof (PBE) generalized gradient approximation (GGA) to describe fully the electron exchange-correlation effect \cite{perdew1996generalized}. Furthermore, we implemented the $GGA+U$ method in the simplified rotationally invariant approach described by Dudarev et al. to address the localized $3d$ electrons in Fe \cite{dudarev1998electron}. The necessity of U$_{eff}$ for Fe was fulfilled by setting $3.7 eV$ based on previous study. we set U$_{eff}$ to be zero for all other elements \cite{liyanage2013theory}. To evaluate the magnetocrystalline anisotropy energy, we first carried out an accurate collinear calculation in the ground state, and then we followed the spin-orbit coupling calculations in two different spin orientations within non-collinear setup.
The substitution of foreign atoms in five crystallographic inequivalent $\mathrm{Fe}$ sites can change the magnetic characteristics of SFO. When foreign atoms are substituted in a SFO unit cell, there are a variety of energetically distinct configurations. The magnetism of substituted SFO is highly dependent on the on-site preferences of the replaced atoms since SFO is ferrimagnetic. Understanding the site preference of substituted atoms is crucial in order to research how substitution affects magnetic characteristics. The substitution energy can be calculated to find the substituted atom's preferred site. The substitution energy $E_{\text {sub}}[i]$ for configuration $i$ at $0\mathrm{~K}$ is given by
\begin{equation}
E_{\mathrm{sub}}[i]=E_{\mathrm{CSFO}}[i]-E_{\mathrm{SFO}}-\sum_{\beta} n_{\beta} \epsilon_{\beta}
\end{equation}
where $E_{\mathrm{CSFO}}[i]$ is the total energy per unit cell of Cr-substituted SFO in configuration $i$, whereas $E_{\mathrm{SFO}}$ is the total energy per unit cell of pure $\mathrm{SFO}$ and $\epsilon_{\beta}$ is the total energy per atom for element $\beta$ ($\beta$= $\mathrm{Cr}$ and $\mathrm{Fe})$ in its most stable crystal structure. $n_{\beta}$ is the number of atoms of type $\beta$ added or removed; if one atom is added, $n_{\beta}=+1$, and when one atom is withdrawn, $n_{\beta}=-1$. The calculation of Magnetic Anisotropy Energy (MAE) is important for understanding the preferred magnetization directions in a material. Mathematically, MAE is defined as the difference between the two total energies where the spin quantization axes are oriented along two distinct directions: \cite{PhysRevB.59.15680}
\begin{equation}
E_{a}=E_{(100)}-E_{(001)}
\end{equation}
where $E_{(100)}$ is the total energy with the spin quantization axis in the magnetically hard axis and $E_{(001)}$ is the total energy with the spin quantization axis in the magnetically easy axis. The total energies in Eq.(2) are computed by the non-self-consistent calculations, where the spin densities are kept constant. 
With the help of MAE, the uniaxial magnetic anisotropy constant, $K_{1}$, can be computed as \cite{munozetal2013, smit1954physical}
\begin{equation}
K_{1}=\frac{E_{a}}{V \sin ^{2} \theta}
\end{equation}
where $V$ is the equilibrium volume of the unit cell and $\theta$ is the angle between the two spin quantization axis orientations $\left(90^{\circ}\right.$ in the present scenario). The anisotropy field, $H_{a}$, which is related to the coercivity can be expressed as \cite{RevModPhys.21.541}
\begin{equation}
H_{a}=\frac{2 K_{1}}{M_{s}}
\end{equation}
where $K_{1}$ is the magnetocrystalline anisotropy constant and $M_{s}$ is the saturation magnetization.

When the difference in substitution energies $E_{\text {sub }}$ between different configurations is relatively small compared to the thermal energy at high annealing temperatures ($\gtrsim 1000 \mathrm{~K})$, the site preference of substituted atoms in hexaferrite can change. This change in site occupation preference can be described using the Maxwell-Boltzmann distribution, which relates to the formation probability. The site occupation probability or the formation probability $P_{i}(T)$ of configuration $i$ at temperature $T$ is given by

\begin{equation}
\begin{aligned}
	P_{i}(T) &=\frac{g_{i} \exp \left(-\Delta G_{i} / k_{B} T\right)}{\sum_{j} g_{j} \exp \left(-\Delta G_{j} / k_{B} T\right)}, \\
 \end{aligned}
 \end{equation}
 \begin{equation}
\Delta G_{i} = \Delta E_{i}+P \Delta V_{i}-T \Delta S_{i}, \\
 \end{equation}
 \begin{equation}
	\Delta S_{i} = k_{B} \ln \left(g_{i}\right)-k_{B} \ln \left(g_{0}\right),
 \end{equation}

where $\Delta G_{i}, \Delta E_{i}, \Delta V_{i}$, and $\Delta S_{i}$ are the change in free energy, substitution energy, unit cell volume, and entropy of the configuration $i$ relative to the ground state configuration. $P$, $k_{\mathrm{B}}$, and $g_{i}$ are the pressure, Boltzmann constant, and multiplicity of configuration i. $g_{0}$ is the multiplicity of the ground state configuration. we considered $\Delta S_{i}$ to be the same for all configurations based on prior literature \cite{dixit2017effect}. Eq.(7) enhances the model through the explicit computation of the entropy change concerning the most stable configuration \cite{gilmore2014materials, RevModPhys.74.11}.

Hence, when the probability of higher energy configurations becomes significant at the annealing temperature, it can be inferred that in a substituted SFO sample, multiple configurations exist rather than a single one. Consequently, any physical quantity of the SFO sample will be a weighted average of the corresponding properties in these different configurations.
\begin{equation}
\langle Q\rangle=\sum_{i} P_{1000 \mathrm{~K}}(i) \cdot Q_{i}
\end{equation}
where $P_{1000 \mathrm{~K}}(i)$ and $Q_{i}$ are the formation probability at $1000\mathrm{~K}$ and a physical quantity $Q$ of the configuration $i$. The weighted average calculated by Eq.(8) is the material’s low-temperature property even though $1000 \mathrm{~K}$
 is used for computation because the crystalline configurations of CSFO will be distributed according to this value during the annealing process.

\section{\label{sec:level3} Results and Discussion}
\indent In order to visualize the doping effect of the $Cr^{+3}$ ion on the structural and magnetic properties of SFO, we replaced the $Fe^{+3}$ from various lattice locations. We found that it has a significant effect on the structural and magnetic properties of this system. We fully relaxed the volume, ionic positions, and shape of SFO and Cr-doped SFO. The crystal structure remains hexagonal under all circumstances. We further carried out our calculations when we match the optimized lattice parameter of pure SFO (a = 5.928 {\AA}, c = 23.195 {\AA}) with experimental lattice constants (a = 5.890 {\AA}, c = 23.182 {\AA}); less than 1$\%$ difference between the calculated and experimental values. For x = 1.0, the calculated lattice parameters (a = 5.930 {\AA}, c = 23.076 {\AA}) were found to be very consistent with experimental lattice parameters (a = 5.902 {\AA}, c = 23.024 {\AA}) \cite{FANG2005281,KIMURA1990186}. Fig.2 shows the variation of lattice parameters from theory and experiment explicitly. 
\begin{figure}[h!]
	 \includegraphics[width=\linewidth]{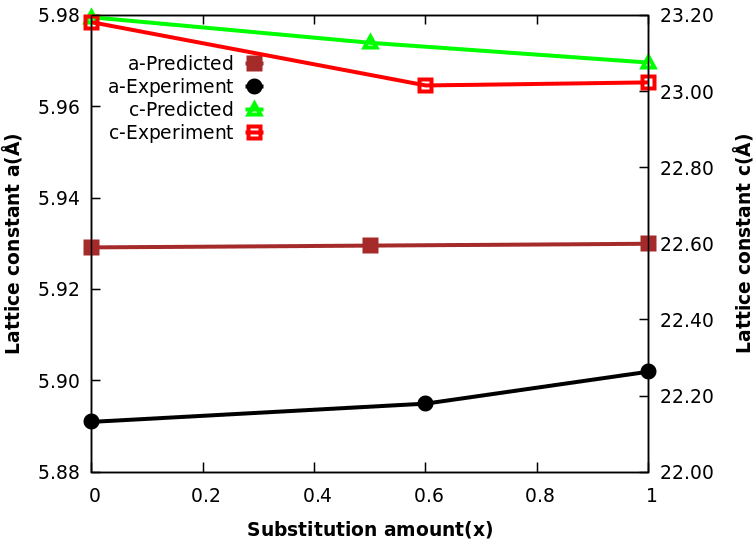}
	\caption{Comparision of predicted and experimental lattice constant of unit cell as a function of fraction of Cr (x).}
\end{figure} 
To compensate for the unavailability of the experimental lattice parameter at x = 0.5, we are comparing the experimental values at x = 0.6 with our calculated values at x = 0.5 for the comparison in Fig. 2. The substitution of Cr in SFO has little effect on the lattice parameters or unit cell volume. Because the radius of Cr$^{+3}$(0.630 {\AA}) is similar to that of Fe$^{+3}$(0.645 {\AA}), this is an expected outcome.

In this paper, we measured the various physical quantities by varying the concentration of Cr in the unit cell of SFO. For the x = 0.5, one Cr atom was substituted at one of the 24 Fe sites of the unit cell. Many of these Fe sites are equivalent when crystallographic symmetry operations are applied, leaving only five inequivalent structures. We label these inequivalent configurations [2a], [2b], [$4f_{1}$], [$4f_{2}$], and [12k] using the crystallographic name of the Fe site. These structures were found by fully optimizing the unit cell shape, volume, and ionic positions. We estimated the substitution energy, $E_{sub}$, to comprehend the site preference of the substituted Cr atom. 

\begin{table*}[!ht]
\caption{\label{tab:table1}Physical properties of inequivalent configurations of $SrFe_{12-x}$Cr$_x$O$_{19}$ with $x = 0$ and $0.5$: Doped amount(x), multiplicity($g$), substitution energy($E_{sub}$),  total magnetic moment ($M_{tot}$), volume of the unit cell ($V$), saturation magnetization ($M_s$), magnetocrystalline anisotropy energy ($E_a$), uniaxial magnetic anisotropy constant ($K_1$), anisotropy field ($H_a$), and the formation probability at $1000$ K ($P_{1000 K}$). All values are for a double formula unit cell containing $64$ atoms.}
\begin{ruledtabular}
\begin{tabular}{ccccccccccc}
        x& config.& g & $E_{sub}(eV)$ & $m_{tot}{(\mu_B)}$ & Volume(\AA$^3$) & $M_s (emu/g)$&$E_a$(meV) & $K_1(KJ/m^3)$ & $H_a(koe)$ & $P_{1000 K}$ \\
        \hline
     0.0& SFO     & -  & -     & 39.99 & 706.18 & 105.19  & 0.87 & 197.12  & 7.50  & 1.000 \\
     0.5&$[2a]$   & 2  & -3.41 & 38.00 & 704.70 & 100.00  & 1.20 & 272.45  & 10.89 & 0.076 \\
	    &$[12k]$  &12  & -3.33 & 38.00 & 705.04 & 100.00  & 0.90 & 204.25  & 8.17  & 0.884 \\
	    &$[4f_2]$ & 4  & -3.27 & 42.00 & 703.70 & 111.00  & 0.90 & 204.63  & 7.39  & 0.049 \\
	    &$[2b]$   & 2  & -2.96 & 38.00 & 703.17 & 100.00  & 0.60 & 136.52  & 5.44  & 0.000 \\
        &$[4f_1]$ & 4  & -2.33 & 42.00 & 705.49 & 111.00  & 0.20 & 45.36   & 1.64  & 0.000 \\
\end{tabular}
\end{ruledtabular}
\end{table*} 

Table I displays the results of our calculation for each of the five inequivalent configurations in ascending order of substitution energy($E_{sub}$). The configuration [2a] has the lowest $E_{sub}$ followed by [12k], and [$4f_{2}$] which is consistent with the experimental outcomes\cite{FANG2005281, parker1980ferrite}. We can conclude that the [2a] site is the most preferred site for the Cr atom at 0 K. We used Eq.(5) to calculate the probability of forming each configuration as a function of temperature.
Because the change in volume between different configurations is so small (less than 0.3 \AA$^3$), we may discard the $P \Delta V$ term as negligible (in the order of $10^{-7} \mathrm{eV}$ at a standard pressure of 1 atm) compared to the $\Delta E_{\text {sub}}$ (i) term in Eq.(6). The entropy change $\Delta S$ has two components: configurational,  $\Delta S_{c}$, and vibrational, $\Delta S_{\mathrm{vib} \cdot}{ }$ \cite{RevModPhys.74.11}. $\Delta S_{\mathrm{vib}}$ is around 0.1-0.2 $k_{\mathrm{B}}$ /atom for binary substitutional alloys like the present system, and $\Delta S_{c}$ is $0.1732 k_{\mathrm{B}}$ /atom. As a result, we assign  $\Delta S = 0.3732 k_{\mathrm{B}} /$ atom.
\begin{figure}[hb]
 	\includegraphics[width=\linewidth]{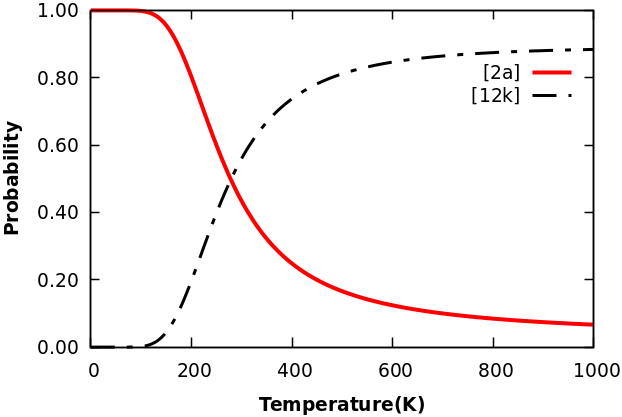}
 	\caption{Temperature dependence of the formation probability of different configurations of $SrFe_{12-x}$Cr$_x$O$_{19}$ with $x = 0.5$. In the plot, the configurations with trivial probabilities are not displayed.}
\end{figure} 

Fig.(3) shows the formation probability of different configurations of $\mathrm{SrFe}_{12-x} \mathrm{Cr}_{x} \mathrm{O}_{19}$ with $x=0.5$ at different temperatures. The foreign $\mathrm{Cr}^{3+}$ ions prefer to occupy host $\mathrm{Fe}^{3+}$ ions from the $2 a$ and the $12 k$ sites. The formation probabilities of $[2b],\left[4 f_{1}\right]$, and $\left[4 f_{2}\right]$ are trivial and not displayed in Fig.(3). A $\mathrm{Cr}^{3+}$ ion has a 100$\%$ probability of occupying the [$2a$] site at 0K and it's value sharply declines as the temperature rises, while the probability of occupancy of $\mathrm{Cr}^{3+}$ at the [$12k$] site is maximum (88 $\%$) at 1000K. At a typical annealing temperature of $1000 \mathrm{~K}$ for $\mathrm{CSFO}$, the site occupation probability of the site $[2a]$ and $[12k]$ are $7\%$ and $88.4\%$ respectively. In CSFO, the doped $\mathrm{Cr}^{3+}$ ions are more likely to occupy $\mathrm{Fe}^{3+}$ ions at the [$12k$] site than the [$2a$] site because of the higher multiplicity of the [12k] site.
\begin{table*}[]
\caption{\label{tab:table2}Physical properties of inequivalent configurations of $SrFe_{12-x}$(Cr)$_x$O$_{19}$ with $x = 1.0$: Doped amount(x), substitution energy($E_{sub}$), total magnetic moment ($m_{tot}$),volume of the unit cell ($V$), saturation magnetization ($M_s$), magnetocrystalline anisotropy energy ($E_a$), uniaxial magnetic anisotropy constant ($K_1$), anisotropy field ($H_a$), and the formation probability at $1000$ K ($P_{1000 K}$). All values are for a double formula unit cell containing $64$ atoms.}
	\begin{ruledtabular}
		\begin{tabular}{ccccccccccc}
		x &config.& $E_{sub}(eV)$ & $m_{tot}{(\mu_B)}$ & Volume(\AA$^3$) & $M_s (emu/g)$ &$E_a$(meV) & $K_1(KJ/m^3)$ & $H_a(koe)$ & $P_{1000 K}$\\
			\hline
1.0&$[2a,2a]$    &  -6.84 & 36.00 & 703.73 & 95.00  & 1.50 & 341.39 & 14.36 & 0.001  \\
  &$[12k,2a]$    &  -6.75 & 30.00 & 703.03 & 79.20  & 1.10 & 250.34 & 12.64 & 0.178  \\
  &$[12k,12k]$   &  -6.69 & 36.00 & 703.24 & 95.00  & 0.80 & 182.01 & 7.66  & 0.664 \\
  &$[2a,4f_2]$   &  -6.68 & 40.00 & 701.63 & 106.00 & 1.30 & 296.45 & 11.20 & 0.009  \\
  &$[12k,4f_2]$  &  -6.62 & 34.00 & 701.89 & 89.70  & 0.90 & 205.16 & 9.13  & 0.145 \\
  &$[4f_2,4f_2]$ &  -6.56 & 44.00 & 700.33 & 116.00 & 1.00 & 228.46 & 7.84  & 0.001 \\   
  &$[12k,2b]$    &  -6.30 & 30.00 & 701.93 & 79.20  & 0.50 & 113.97 & 5.75  & 0.001 \\
  &$[2b,4f_2]$   &  -6.15 & 34.00 & 698.58 & 89.70  & 0.70 & 160.33 & 7.10  & 0.000  \\
  &$[2a,2b]$    &  -6.09 & 30.00 & 701.90 & 79.20  & 0.40 & 91.18  & 4.60  & 0.000  \\
  &$[2b,2b]$    &  -5.99 & 36.00 & 699.37 & 95.00  & 2.00 & 457.55 & 19.15 & 0.000  \\
  &$[12k,4f_1]$ &  -5.66 & 40.00 & 704.40 & 106.00 & 0.20 & 45.43  & 1.72  & 0.000  \\
  &$[4f_1,4f_2]$ &  -5.61 & 44.00 & 703.98 & 116.00 & 0.20 & 45.46  & 1.57  & 0.000  \\
  &$[2a,4f_1]$  &  -5.47 & 40.00 & 704.37 & 106.00 & 0.90 & 204.44 & 7.76  & 0.000  \\
  &$[2b,4f_1]$  &  -5.15 & 38.91 & 702.14 & 103.00 & 0.10 & 22.79  & 0.88  & 0.000 \\
  &$[4f_1,4f_1]$ & -4.67 & 44.00 & 705.71 & 116.00 & 0.60 & 136.03 & 4.70  & 0.000 \\      
    \end{tabular}
	\end{ruledtabular}
\end{table*}

\begin{table*}[hbt!]
 \caption{\label{tab:table4}Weighted averages of physical properties of pure and $\mathrm{Cr}$-doped strontium hexaferrite: volume of the unit cell $(\mathrm{V})$, total magnetic moment $\left(m_{\text {tot }}\right)$, saturation magnetization $\left(M_{s}\right)$, magnetocrystalline anisotropy energy ($\left.E_{a}\right)$), uniaxial magnetic anisotropy constant $\left(K_{1}\right)$, and anisotropy field $\left(H_{a}\right)$. The column labeled $\left (M_{s}\right)$ displays the calculated values at 0K along with the bracketed experimental values from K. Praveena et al. \cite{praveena2014} for the comparison.}
\begin{tabular}{lcccccc}
 	\hline\hline

 Material & $V$ (\AA$^3$) & $m_{\text{tot}}$ ($\mu_{\mathrm{B}}$)& $M_{s}$ (emu/g)& $E_{a}$ (meV)& $K_{1}$ (kJ/m$^{3}$)& $H_{a}$ (kOe) \\
 	\hline
 	$\mathrm{SrFe}_{12} \mathrm{O}_{19}$ & $706.18$ & $39.99$ & $105.19 (59.33)$ & $0.87$ & $197.12$ & $7.50$ \\
 	$\mathrm{SrFe}_{11.5} \mathrm{Cr}_{0.5} \mathrm{O}_{19}$ & $704.39$ & $38.16$ & $100.46 (36.01)$ & $0.91$ & $208.58$ & $8.30$ \\
 	$\mathrm{SrFe}_{11.0} \mathrm{Cr}_{1.0} \mathrm{O}_{19}$ & $702.07$ & $34.63$ & $91.40 (30.04)$ & $0.87$ & $198.44$ & $8.79$ \\
 	\hline\hline
   \end{tabular}
\end{table*}
For the x = 1.0, two Cr atoms were substituted at two of the 24 Fe sites of the unit cell. Many of these Fe sites are equivalent when crystallographic symmetry operations are applied, leaving only 15 inequivalent structures. These structures were found by fully optimizing the unit cell shape, volume, and ionic positions. To understand the site preference of a substituted Cr atom, we estimated the substitution energy, $E_{sub}$.
\begin{figure}[]
	\includegraphics[width=\linewidth]{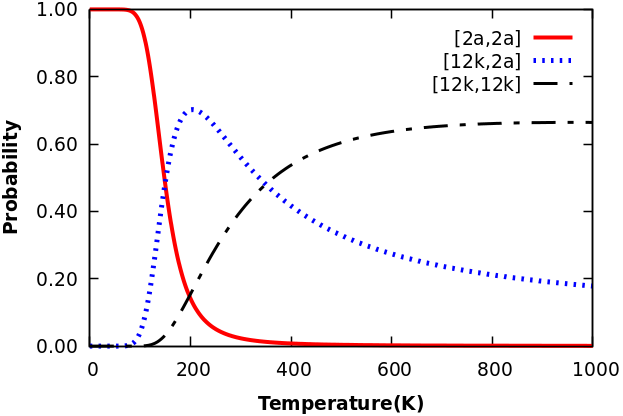}
	\caption{Temperature dependence of the formation probability of different configurations of $SrFe_{12-x}$Cr$_x$O$_{19}$ with $x = 1.0$. In the plot, the configurations with trivial probabilities are not displayed.}
\end{figure} 
Table II provides the results of our calculation for each of the fifteen inequivalent configurations in ascending order of substitution energy($E_{sub}$). The configuration [2a, 2a] has the lowest $E_{sub}$ followed by [12k, 2a], and [12k, 12k].  We also used Eq.(5) to calculate the probability of forming each configuration as a function of temperature. Because the change in volume between different configurations is so small (less than 0.7\AA$^3$), we may discard the $P \Delta V$ term as negligible (in the order of $10^{-7} \mathrm{eV}$ at a standard pressure of 1 atm) compared to the $\Delta E_{\text {sub }}$ (i) term in Eq.(6). 

Fig. (4) shows the formation probability of different configurations of $\mathrm{SrFe}_{12-x} \mathrm{Cr}_{x} \mathrm{O}_{19}$ with $x=1.0$ at different temperatures. The foreign $\mathrm{Cr}^{3+}$ ions prefer to occupy host $\mathrm{Fe}^{3+}$ ions from the $[2a, 2a]$, $[12k, 2a]$,and $[12k, 12k]$ sites. The formation probabilities of other sites are trivial and not displayed in Fig.(4). A $\mathrm{Cr}^{3+}$ ion has a 100$\%$ probability of occupying the $[2a, 2a]$ site at 0K and it's value sharply declines as the temperature rises, while the probability of occupancy of $\mathrm{Cr}^{3+}$ at the $[12k, 12k]$ site is maximum (66.4 $\%$) at 1000 K. At a typical annealing temperature of $1000 \mathrm{~K}$ for $\mathrm{CSFO}$, the site occupation probability of the site $[12k, 2a]$ is $17.8\%$. In CSFO, the doped $\mathrm{Cr}^{3+}$ ions are more likely to occupy $\mathrm{Fe}^{3+}$ ions at the $[12k, 12k]$ site than the $[2a, 2a]$ site because of it's higher multiplicity. We exclusively utilize the formation probability at elevated temperatures for computing weighted averages. This is because the arrangement of CSFO configurations during the annealing process will be distributed according to these values. Table III displays the weighted average of corresponding quantities as the concentration of $\mathrm{Cr}^{3+}$ increases. The volume of CSFO decreases as we increase the concentration of $\mathrm{Cr}^{3+}$ because of the smaller atomic radius of $\mathrm{Cr}^{3+}$ ion. The magnetic moment of CSFO is also in the decreasing trend as the amount of doped $\mathrm{Cr}^{3+}$ increases owing to its smaller magnetic moment (3{$\mu_B$}) than the $\mathrm{Fe}^{3+}$ (5{$\mu_B$}). Similarly, the saturation magnetization is decreasing monotonically as we increase the $\mathrm{Cr}^{3+}$ concentration which is consistent with the K. Praveena et al.\cite{praveena2014}. Although the value of magnetocrystalline anisotropy $\left(K_{1}\right)$ is slightly increased, the reduction in saturation magnetization $\left (M_{s}\right)$ is much more significant. So, their resultant effect causes the anisotropy field $\left(H_{a}\right)$ to increase as the fraction of Cr is raised. In Table IV, we have provided the atomic contribution from each sublattice to the overall magnetic moment of CSFO. It can be observed that the total magnetic moment of the unit cell is slightly distinct from the sum of local magnetic moments. This disparity arises from the contribution of the interstitial region to the overall magnetic moment.
\begin{table}[htbp]
\scriptsize
\caption{\label{tab:table5}Contribution of atoms in each sublattice to the total magnetic moment of $\mathrm{Cr}$-substituted strontium hexaferrite structures [ $12 \mathrm{k}]$ and $[12k, 12k]$ compared with pure SFO. All magnetic moments are in $\mu_{B}$, and $\Delta m$ is measured relative to the values for pure SFO.}
\begin{tabular}{ccccccccccc}
\hline\hline
		& \multicolumn{2}{c}{SFO} &       & \multicolumn{3}{c}{x = 0.5, [12 k]} &       & \multicolumn{3}{c}{x = 1.0, [12 k, 12 k]} \\
		\cline{2-3}\cline{5-7}\cline{9-11}    
		Site  & Atoms & m     &       & Atoms & m     & $\Delta{m}$ &       & Atoms & m     & $\Delta{m}$ \\
		\hline
	    2d     & 2 Sr  & -0.006 & &2 Sr&-0.005  & 0.001&   & 2 Sr  & -0.005 & 0.001 \\
        2a     & 2 Fe  &  8.328 & &2 Fe& 8.334 & 0.006 &   & 2 Fe  & 8.336  & 0.008 \\        
        2b     & 2 Fe  &  8.130 & &2 Fe& 8.131 & 0.001 &   & 2 Fe  & 8.129  & -0.001 \\
      $4f_{1}$ & 4 Fe  & -16.184& &4 Fe&-16.159 & 0.025&   & 4 Fe  &-16.141 & 0.043 \\
      $4f_{2}$ & 4 Fe  & -16.416& &4 Fe&-16.410 & 0.006&   & 4 Fe  &-16.415 & 0.001 \\
      12k      & 1 Fe  & 4.180  & &1 Cr& 2.680  & 0.319&   & 1 Cr  & 2.689  & 0.311 \\
               & 1 Fe  & 4.178  & &1 Fe& 4.180  & 0.000 &  & 1 Cr  & 2.692  & 0.308 \\
               & 10 Fe & 41.794 & &10 Fe& 41.792 & -0.002& & 10 Fe & 41.805 & 0.011 \\
      4e       & 4 O   & 1.400  & &4 O & 1.292   & -0.108& & 4 O   & 1.165  & -0.235 \\
      4f       & 4 O   & 0.356  & &4 O & 0.294   & -0.062& & 4 O   & 0.239  & -0.117 \\
      6h       & 6 O   & 0.160  & &6 O & 0.151   & -0.009& & 6 O   & 0.142  & -0.018 \\      12k      & 24 O  & 3.140  & &24 O& 2.777   & -0.363& & 24 O  & 2.422  & -0.718 \\
   $\Sigma{m}$ &       & 39.053 & &    & 37.056  &       & &       & 35.057 &       \\
    $m_{tot}$  &       & 40     & &    &   38    &       & &       &   36   &       \\
\hline\hline
\end{tabular}
\label{tab:addlabel}
\end{table}
\section{\label{sec:level4} Conclusions}
First-principles total-energy calculations based on density functional theory were used to study Cr-substituted SFO ($\mathrm{SrFe}_{12-x} \mathrm{Cr}_{x} \mathrm{O}_{19}$) with $x = 0.0, 0.5, 1.0$. The results showed that increasing the fraction of Cr atoms reduced the total magnetic moment of the SFO unit cell. This reduction in magnetization was obtained by low magnetic Cr atoms replacing $\mathrm{Fe}^{3+}$ ions at two of the majority spin sites, 2a and 12k, resulting negative contribution to the magnetization. Our substitution energy and formation probability analysis predicts that Cr atoms preferentially occupy the 2a, 12k, and 4f$_{2}$ sites, consistent with experimental observations. Increasing the fraction of Cr in SFO leads to a rise in the magnetic anisotropic field $\left(H_{a}\right)$ despite a decrease in magnetization and a slight increase in magnetocrystalline anisotropy $\left(K_{1}\right)$. This increase in anisotropic field$\left(H_{a}\right)$ is supported by a formation probability study, which shows that at higher temperatures (>700°C), $\mathrm{Cr}^{3+}$ ions are more likely to occupy the 12k site rather than the 2a site because of its higher multiplicity.

\begin{acknowledgments}
This work was supported by the Center for Computational Science (CCS) at Mississippi State University. Computer time allocation has been provided by the High-Performance Computing Collaboratory $(HPC^{2})$ at Mississippi State University.
\end{acknowledgments}

		\nocite{*}
		\bibliography{mybib}    
\end{document}